\documentclass[10pt]{article}
\usepackage[utf8]{inputenc}
\usepackage{amsmath,amsfonts,amssymb,latexsym}
\usepackage[T1]{fontenc}
\newtheorem{satz}{Theorem}[section]

\newtheorem{defi}[satz]{Definition}

\newtheorem{bem}[satz]{Remark}
\newtheorem{lemma}[satz]{Lemma}
\newtheorem{koro}[satz]{Corollary}
\newtheorem{conclusion}[satz]{Conclusion}
\newtheorem{ob}[satz]{Observation}

\newtheorem{propo}[satz]{Proposition}

\newtheorem{conjecture}[satz]{Conjecture}

\newcommand{\mcal}{\mathcal}
\newcommand{\mbf}{\mathbf}

\newcommand{\tit}{\textit}

\newcommand{\C}{\mathbb{C}}

\newcommand{\R}{\mathbb{R}}

\newcommand{\beq}{\begin{equation}}
\newcommand{\eeq}{\end{equation}}

\newcommand{\cA}{\mcal{A}}
\newcommand{\cH}{\mcal{H}}
\newcommand{\cB}{\mcal{B}}

\begin{document}
\thispagestyle{empty}
\begin{center}
\vspace*{1.0cm}
{\Large{\bf The Role of Type $II_{\infty}$ v.Neumann Algebras\\ and their Tensor Structure
in Quantum Gravity}}
\vskip 1.5cm
 
{\large{\bf Manfred Requardt}}

\vskip 0.5cm

Institut fuer Theoretische Physik\\
Universitaet Goettingen\\
Friedrich-Hund-Platz 1\\
37077 Goettingen \quad Germany\\
(E-mail: muw.requardt@googlemail.com) 

\end{center}
\begin{abstract}
We will argue in this paper that the type classification of v.Neumann algebras play an important role in a theory of quantum gravity and quantum space-time physics. We provide arguments that type $II_{\infty}$ and its representation as a tensor product of an ordinary (exterior) Hilbert space algebra $\cB(\cH_I)$ and an (internal) type $II_1$ algebra, encoding, in our view, the hidden microscopic gravitational degrees of freedom, do represent the first step away from the semiclassical picture towards a full theory of quantum gravity.
\end{abstract}
\newpage
\section{Introduction}
We want to show that the type classification of v.Neumann algebras is playing an important role in a theory of quantum gravity (QG). This type classification was developed in four fundamental papers by v.Neumann and Murray and v.Neumann in the nineteen thirties/fourties of the past century (\cite{N1},\cite{N2},\cite{N3},\cite{N4} or see \cite{N5}).

While v.Neumann considered the type two to be the most promising (for various reasons) it played only a minor role in modern mathematical physics with type one and three being of major importance in relativistic quantum field theory (RQFT) and quantum statistical mechanics of infinitely extended systems. We wiil however argue that the type two encodes fundamental structural aspects of a future theory of QG.

We want to briefly sketch why, in our view, this is the case. We think, the main point is the way and extent how microscopic information is distributed in the quantum systems under investigation and after all in quantum space-time (QST) itself and its availability via measurements and observations. For example, in ordinary non-relativistic quantum theory complete information (at least in principle) is available. Therefore the theory is usually formulated over \tit{pure quantum states}. Correspondingly, the v.Neumann algebra of observables contains the projectors on pure states and is of type one.
\begin{bem}Note that in quantum theory a crucial part is the preparation and reduction by measurements of states.
\end{bem}

In RQFT the situation is different in so far as the availability of full information by local observations is restricted and prohibited by the existence of the finite velocity of light (causality) and the commutativizy of space-like local observables (locality) which, furthermore results in the non-existence of localized pure states and local projections on such pure states. The corresponding v.Neumann algebras of observables are typically of type III.

We now come to QG, a theory which does not really exist at the moment but some aspects of a possible future framework are, in our view, just emerging. Recently there appeared several papers by Witten and coworkers, dealing with various aspects of the \tit{holographic principle} and \tit{black holes}, in which, apart from the more ordinary type III types of observable algebras also the type II classification played a certain role (\cite{W1},\cite{W2},\cite{W3},\cite{W4}). 

We can characterize the different possible types of v.Neumann algebras in the following way:
\begin{satz}Type I is characterized by the existence of minimal or abelian projectors  (e.g. pure states). In type $II_1$ all projectors are finite, in type $II_{\infty}$ there exist both finite and infinite projectors. Furthermore, and most importantly, in type II there exist no minimal projectors. In type III, a case we do not discuss in this paper, all projectors are infinite. The technical details are given in the next section. Physically, instead of pure states we have at least mixtures and density matrices in the case $II_1$, i.e., a (quasi) thermal behavior.
\end{satz}
\begin{bem}One should note that these properties are of a higly technical character and are only indirectly corresponding to physical properties.
\end{bem}

We want to argue in this paper that the generic observable algebras in QG are of type $II_{\infty}$ which turn out to be unitarily equivalent to a v.Neumann tensor product of an ordinary $\mcal{B}(\mcal{H}_I)$ (the full algebra of bounded operators on an infinite dimensional, separable Hilbert space) with a type $II_1$ factor, which, in our view, can be envisaged to represent the quantum mechanical fine structure of the microscopic gravitational degrees of freedom (DoF) in the infinitesimal microscopic neighborhood of the respective macroscopic points of our smooth space-time manifold ST.

We want to conclude this section with the following remark:
\begin{ob}In addition to what we have said in the preceding paragraphs, we want to make the following comment. One of our main observations is that, in our view, the type classification in the QG context tells us something about the kind of interaction which prevails among the microscopic DoF of QST. It implies for example that in QST only statistical mixtures and density matrices can be measured and/or observed locally, a reason being that the infinitesimal neighborhoods of the macroscopic points  $x\in ST$ are essentially open systems with respect to their environment in QST.
\end{ob}

In the following section we sum up some relevant results about v.Neumann algebras and their type classification in so far as they are of importance in our analysis, which is given in section three. We decided to not transferring this material to an appendix because it will be essential for the understanding of what we are doing in section three. 
\section{A brief Outline of v.Neumann Algebras and their Type Classification}
The type classification of v.Neumann algebras was developed by Murray and v.Neumann. It was a deep observation that v.Neumann algebras were not necessarily of the type of a $\mcal{B}(\mcal{H})$, (i.e., the bounded operators over some separable Hilbert space). It turns out that their type crucially depends on the character of the projectors they contain and which generate them. In the following we denote v.Neumann algebras by $\mcal{A}$, projectors by $P_j$, Hilbert spaces by $\mcal{H}_i$, unitary operators by $U_j$, partial isometries by $V_{ij}$ with
\begin{defi}A partial isometry maps bijectively and isometrically 
\beq V_{ij}:\; P_i\mcal{H}\;\to\; P_j\mcal{H}  \eeq
It holds:
\beq V_{ij}^*\cdot V_{ij}=P_i\; ,\; V_{ij}\cdot V_{ij}^*=P_j  \eeq
\end{defi}
\begin{bem}The projectors of a v.Neumann algebra form a lattice with respect to the partial order $\leq$. It holds
\beq P_i\leq P_j\;\text{if}\; P_i\mcal{H}_i\subseteq P_j\mcal{H}_j  \eeq
\end{bem}

A v.Neumann algebra $\mcal{A}$ is defined by v.Neumann as a $\ast$-subalgebra of some $\mcal{B}(\mcal{H})$ (together with $A$, $A^*$ is an element of $\mcal{A}$) with
\beq   \mcal{A}=\mcal{A}''  \eeq
where $\mcal{A}'$ denotes the commutant of $\mcal{A}$. It turns out that a v.Neumann algebra is closed in the weak and strong (and some other) topologies.
\begin{defi}$\mcal{A}$ is weakly, strongly closed if for all $x,y\in\mcal{H}$ it holds:
\beq (x,A_n y)\to (x,Ay)\quad \text{for all} \quad x,y \eeq
or
\beq A_n x\to Ax\quad\text{in norm}  \eeq
implies that $A\in\mcal{A}$.
\end{defi}
\begin{defi}With $S$ a subset of $\mcal{A}$ we say, $S$ generates $\mcal{A}$ if $S''=\mcal{A}$
\end{defi}
\begin{lemma}Both the projectors and the unitaries generate $\mcal{A}$.
\end{lemma}

It is interesting that v.Neumann algebras can be defined in an equivalent but different way.
\begin{propo}Let $\mcal{A}$ be a $\ast$-algebra with unit $\mbf{1}$, then it holds for the weak, strong closures of $\mcal{A}$: $\mcal{A}_w=\mcal{A}_s=\mcal{A}''$.
\end{propo}
Crucial is the property that with
\beq A_n\to A\;,\;B_n\to B\quad\text{strongly}\quad  \eeq
\beq A_nB_n\to AB\quad\text{strongly}  \eeq
as we have
\beq A_nB_n-AB=A_n(B_n-B) +(A_n-A)B \eeq
and $||A_n||$ uniformly bounded for weak or strong operator convergence. We hence see that the strong closure is again a star algebra.
\begin{bem}For the weak closure and weak convergence the above construction does not work.
\end{bem}
 (For proofs see e.g. \cite{Dix} corollary 1 on p.45 and the rather technical proof of Th.11 on p.54 in \cite{Top}). For the weak closure one shows instead that weak and strong closure are the same which is far from evident.
\begin{bem}For convenience we treat in the following only socalled factors. For factors it holds
\beq  \cA\cap\cA'=\C\cdot\mbf{1}  \eeq
\end{bem}

Before we begin to give a brief description of the type classification of v.Neumann algebras, we want to mention some of the relevant literature. We have to cite a number of sources as they vary greatly in character and not every important aspect of the theory will be found in each of the references. 
\begin{bem}Note that we are primarily interested in the case of type II.  \end{bem}
Older and more recent books are for example \cite{Dix},\cite{Kad2},\cite{Jones},\cite{Tak1},\cite{Black},\cite{Sak},\cite{Top},\cite{Strat},\cite{Sund}. A, in our view, extremely well-written paper is \cite{Sorce}, in which it is tried to relate the more technical results of the theory to various more physical aspects and to elucidate their relevance.

In addition to the partial order, given by $\leq$ among projections, we now introduce another one denoted by $\preceq,\succeq,\prec,\succ,\sim$.
\begin{defi}With $P,Q$ two projections in $\cH$ we say $P\sim Q$ if there exists a partial isometry
\beq U_{PQ}:\; P\cH \to Q\cH\,\text{with}\, U_{PQ}^*\cdot U_{PQ}=P\,\text{,} \,U_{PQ}\cdot U_{PQ}^*=Q \eeq
If there exists a $Q_1\leq Q$ with $P\sim Q_1$ we write $P\preceq Q$. We use the symbol $\prec$ if $Q_1$ is a true subprojection of $Q$, i.e. $Q_1\cH\subset Q\cH$.
\end{defi}
\begin{propo}$\preceq$ is both reflexive and transitive, i.e.
\beq P\preceq Q\;\text{and}\; Q\preceq P\;\text{implies}\; P\sim Q  \eeq
and
\beq P\preceq Q\;\text{and}\; Q\preceq L\;\text{implies}\; P\preceq L  \eeq
\end{propo}
\begin{bem}In this context, as is the case for many of these seemingly plausible results, the proofs are frequently quite intricate.
\end{bem}
We have the following important result:
\begin{satz}If $\cA$ is a factor, it holds for all $P,Q$ that we have either
\beq P\preceq Q\;\text{or}\; Q\preceq P  \eeq
\end{satz}
that is, we have a total order.

We now come to a more detailed characterization of projectors and v.Neumann algebras.
\begin{defi}i)A projector, $P$, is called minimal in $\cA$ if there does not exist a true subprojection $Q\in\cA$ with $Q<P$.\\ 
ii)A projector is called finite if there does not exist a true subprojection $Q<P$ with $Q\sim P$.\\
iii)A projector $P$ is called infinite if it is not finite, i.e., it exists a true subprojection $Q<P$ with $Q\sim P$.\\
iv)A projector $P$ is called purely infinite if no subprojection is finite.
\end{defi}
\begin{bem}In some definitions the existence of a single finite or infinite projector is assumed in $\cA$, but we will show immediately that this usually implies that many projectors of this kind will exist  (see below). Hence, the existence of many projectors of the respective type is the true characteristic of the v.Neumann algebras under discussion.
\end{bem}

\begin{satz}(Structure Theorem) i)A v.Neumann factor is called to be of type I if it contains minimal projectors.
ii)It is called to be of type III if it does not contain finite projectors (a minimal projector is finite).
iii)It is of type II if it does not contain minimal projectors and more specifically, it is of type $II_1$ if all projectors are finite and of type $II_{\infty}$ if it contains finite and infinite projectors.
\end{satz}
\begin{propo}If $E$ is a finite projection, each subprojection is finite. By the same token, if $E$ is infinite, each $Q\geq E$ is infinite.
\end{propo}
Proof: Assume $E_0<E$ and $E_1<E_0$ with $E_1\sim E_0$. With $V$ the partial isometry from $E_0$ to $E_1$ we have
\beq E-E_0+E_1\sim E-E_0+E_0=E\;\text{and}\; E-E_0+E_1< E  \eeq
which is a contradiction.

Frequently  projectors and/or v.Neumann algebras are characterized by the type of the unit projector $\mbf{1}$ and it is said that this is equivalent to the preceding type characterizations. We want to prove the following:
\begin{satz}i)If $\mbf{1}$ is finite the same holds for all subprojections in $\cA$.\\
ii)If $\mbf{1}$ is infinite, there exists a projector $P$ in $\cA$ and in fact a hole sequence which is infinite.
\end{satz}
Proof: The first part follows from the above proposition. On the other hand, let $\mbf{1}$ be infinite. hence, there exists a true subprojection $P$ with $P\sim\mbf{1}$ and a partial isometry 
\beq V:\cH\to P\cH\;,\; VV^*=P\;,\; V^*V=\mbf{1}  \eeq
We have
\beq V^2:\cH\to P\cH\to Q\cH\quad\text{with}\quad Q<P  \eeq
\beq (V^*)^2:Q\cH\to\cH\;,\;(V^*)^2V^2=\mbf{1}\;,\; V^2(V^*)^2= Q  \eeq
more specifically
\beq VP:P\cH\to Q\cH\;,\;PV^*VP=P\;,\; VP^2V^*=Q  \eeq
Hence it holds that $P\sim Q$ with $Q<P$, i.e. $P$ is infinite if $\mbf{1}$ is infinite. Furthermore, we have that $\mbf{1}$ is infinite if $P$ is infinite. That is, we have the full equivalence of the respective statements.\\
$P$ finite, infinite and $\mbf{1}$ finite, infinite.
\begin{koro}We see that by this construction we can generate a whole sequence of infinite projectors
\beq \mbf{1}>P_1>P_2\cdots  \eeq
with the help of partial isometries $V^n,(V^*)^n$.
\end{koro}

From a more physical point of view the relation between the two partial orders $\leq\;,\;\preceq$ is of interest.
\begin{koro}An infinite projector has infinite range dimension. In a type II v.Neumann algebra also a finite projector has infinite range dimension.
\end{koro}
Proof:If $Q<P$ and $Q\sim P$ the range dimension of $P$ cannot be finite as the partial isometry restricted to $V:Q\cH\to P\cH$ is an isometry but due to $Q<P$ we have $dim(Q\cH)< dim(P\cH)$ which cannot hold.\\
On the other hand, in a type II v.Neumann algebra there are no minimal projectors. That is, for each $P$ there exists a $Q$ in $\mcal{A}$ with $Q<P$. If $dim(P\cH)=n<\infty$ it follows that $dim(Q\cH)<n$. i.e., in a finite number of steps we arrive at one-dimensional projectors which are minimal. This is a contradiction, hence $dim(P\cH)=\infty$.
\section{The Tensor Product Structure of Type $II_{\infty}$ v.Neumann Algebras and its Implications for Quantum Gravity and Quantum Space-Time}
Our strategy in this section is the following. There do exist various approaches in QG which try to reconstruct the (quantum) physics on the more ordinary scales of energy and space or time by starting from the very remote Planck scale and, working from the bottom up, hope to reach with the help of a lot of speculation and intermediary steps the ordinary level of known and well established physics.

We want to persue a different more top down strategy. We will argue that the tensor structure of type $II_{\infty}$ v.Neumann algebras represents the first (quantum) step away from the usual semiclassical framework, that is, QFT on curved space-time.
\subsection{The Tensor Product Structure of Type $II_{\infty}$ v.Neumann Algebras}
We argued in the introduction that type I is typical for non-relativistic quantum mechanics and type III occurs primarily in quantum statistical mechanics in infinite systems and quantum field theory. In this section we want to explain why we think that type $II_{\infty}$ and type $II_1$ are relevant for an approach to QG.
 
 In a first step we will prove that type $II_{\infty}$ v.Neumann algebras have a natural tensor product structure.
 \begin{satz}i)In a type $II_{\infty}$ v.Neumann algebra there exists an infinite family of mutually orthogonal finite projectors $\{P_i\}$ with all $P_i\sim P_0$ and $\sum\, P_i=1$.
 ii)A type $II_{\infty}$ v.Neumann algebra is unitarily equivalent to a tensor product of the form
 \beq \cB(\cH_I)\,\overline{\otimes}\,P_0\cA P_0  \eeq
$P_0$ a finite projector in $\cA$ (acting on $\cH$) and $\overline{\otimes}$ the canonical v.Neumann tensor product (the double commutant or weak/strong closure of the algebraic tensor product). $\cH_I$ is a separable Hilbert space with ON basis labelled by a countable index set $I$(see below):
 \end{satz}
 \begin{bem}In most sources the unitary equivalence as such is emphasized. But we know that all separable Hilbert spaces are unitarily equivalent and hence unitarily equivalent operator algebras do exist. From a physical point of view what is actually important is the way how the relevant observables are expressed in the representations under discussion. This can happen in a very transparent or, on the other side, highly intransparent way. Therefore in the following proof we are also interested in the way the observables in $\cA$ are represented (cf. e.g. \cite{Strat}).
 \end{bem}
 Proof: In a type $II_{\infty}$ algebra we have a finite projector, denoted by $P_0$. By induction we choose a maximal family of pairwise orthogonal projectors $\{P_i\}$, each $\sim P_0$. If $\sum\, P_i\neq \mbf{1}$ we have by the comparison theorems for factors that either
 \beq \mbf{1}-\sum\, P_i\preceq P_0\;\text{or}\; P_0\preceq \mbf{1}-\sum\, P_i  \eeq
The latter case would imply that $\{P_i\}$ is not maximal, hence we have
\beq \mbf{1}-\sum\, P_i\preceq P_0  \eeq
It follows that $\mbf{1}-\sum\, P_i$ is finite. If the above sum is finite it follows that $\sum\, P_i$ is also finite, hence $\mbf{1}$ is finite in a type $II_{\infty}$ algebra, which is a contradiction. That is, the above sum is infinite, i.e. we have $\sum_0^{\infty} P_i$.
 
 We now show that $\sum_0^{\infty} P_i=\mbf{1}$. We have
 \beq \mbf{1}=\sum_0^{\infty}\,P_i+(\mbf{1}-\sum_0^{\infty}\,P_i)  \eeq
 Writing $\sum_0^{\infty}P_i\;\text{as}\; (P_0+\sum_1^{\infty}P_i) $
 and using the `'bijection trick'' (shift of indices)
\beq U: \sum_0^{\infty}\,P_i\,\to\,\sum_1^{\infty}\,P_i\; ,\;P_i\to P_{i+1}  \eeq
(which works if the sum is infinite), yielding $\sum_0^{\infty}\,P_i\sim\sum_1^{\infty}\,P_i$, we get finally
\beq \mbf{1}=(\mbf{1}-\sum_0^{\infty}\,P_i)+\sum_0^{\infty}P_i\prec P_0+\sum_1^{\infty}P_i=\sum_0^{\infty}P_i\prec\mbf{1} \eeq
This is a contradiction, hence we have $\mbf{1}=\sum\, P_i$.

It remains to prove the second statement, i.e., the unitary equivalence of $\cA$ and 
$\cB(\cH_I)\,\overline{\otimes}\,P_0\cH$. We choose an ON basis $\{e_{i0}\}$ in $P_0\cH$.
Using the partial isometries
\beq V_j:P_0\to P_j  \eeq
we get ON bases in $P_j\cH$ via $V_je_{i0}:= e_{ij}$, which yields an ON basis in $\cH$.
\beq \cH\ni x=\sum\, x_{ij}e_{ij} \eeq
With $V_j^*e_{ij}=e_{i0}$ we define an unitary map
\beq U:e_{ij}\mapsto e_{i0}\otimes \psi_j\in P_0\cH\otimes\cH_I  \eeq
$\{\psi_j\}$ an ON basis in some Hilbert space which is labelled by the indices of the projectors $\{P_j\}$.

This map is unitary as we have 
\beq ||Ux||^2=||\sum\, x_{ij}\cdot e_{i0}\otimes\psi_j||^2=\sum\, |x_{ij}|^2 \eeq
We now have (with $T\in\cA$):
\beq UTU^{-1}(e_{i0}\otimes\psi_j)=UTe_{ij}=U(\sum_{i'j'}(e_{i'j'}|Te_{ij})\cdot e_{i'j'} \eeq
which equals
\beq \sum_{i'j'}(e_{i'j'}|Te_{ij})\cdot e_{i'0}\otimes\psi_{j'}  \eeq
That is, $(e_{i'j'}|Te_{ij})$ is the matrix element belonging to\\ $(e_{i'0}\otimes\psi_{j'}|(UTU^{-1})e_{i0}\otimes\psi_j)$.
It can now be easily inferred that 
\beq U\cA U^{-1}=P_0\cA P_0\overline{\otimes}\cB(\cH_I)  \eeq
as bounded operators are uniquely given by their matrix elements and the above relation can be run in both ways, from $\cH$ to $P_0\cH\otimes\cH_I$ or the other way around.

It may be useful to give an explicit representation of the operator $UTU^{-1}$ in $P_0\cA P_0\overline{\otimes}\cB(\cH_I)$. It reads:
\beq UTU^{-1}=\sum_{i,k\in I}(V_k^*TV_i)\otimes W_{ki}  \eeq
which shows that it is really an operator in $P_0\cA P_0\overline{\otimes}\cB(\cH_I)$. The $W_{ik}$ are shift operators in $\cB(\cH_I)$. They are partial isometries
\beq W_{ik}:[\C\cdot\psi_i]\to[\C\cdot\psi_k]   \eeq
and generate $\cB(\cH_I)$
\begin{bem} They are playing the role of socalled matrix units (cf. e.g. \cite{Kad2}).
\end{bem}
\begin{bem} Note, that in $\cA$ acting on $\cH$ we have
\beq \cA=\mbf{1}\cdot\cA\cdot\mbf{1}=\sum_{ij}\, P_j\cdot \cA\cdot P_i  \eeq
\end{bem}
\subsection{Physical Implications for Quantum Gravity and Quantum Space-time}
In a certain sense one can consider the representation $\cA_{(II_{\infty})}\cong P_0\cA P_0\overline{\otimes}\cB(\cH_I)$ with $P_0\cA P_0$ of type $II_1$ as sort of a bundle structure (unfortunately it does not fulfill all the bundle axioms). We can regard $\cH_I$ as kind of an exterior or base space with an internal space, $P_0\cH$, or $P_i\cH$ being attached to each coordinate parameter (or base point) $i\in I$ of the exterior space. 

It is interesting to relate this structure of an internal space of type $II_1$ attached to each point $i\in I$ to the corresponding structure of an algebra of type $I_n$ or $I_{\infty}$. In the latter situation (as is emphasized in \cite{Sorce}) one has an internal space without any structure. Instead of  a full algebra of type $II_1$ there occurs only the unit projector $\mbf{1}$ in the corrsponding tensor product representation. On the other hand, we have seen that the finite projectors in a type $II_1$ algebra have a quite complex structure as there do not exist finite rank projectors. With $P_1$ finite there always exists (as $P_1$ is not minimal) another finite projector $P_2<P_1$ and so on. 
\begin{ob}As in quantum theory projectors, $P_i$, and their corresponding subspaces, $P_i\cH$, are closely related to preparation of states and possible observations, the above nested inclusion of classes of projections may hint at the existence of consecutive layers of coarse graining and/or scale dependent resolution of (quantum) space-time ((Q)ST). 
\end{ob}

A central role in ordinary (non-relativistic) quantum mechanics is played by the notion of state, in particular, pure states, from which the mixtures, i.e., density matrices are built. Algebraically a state is a positive, linear functional
\beq \omega:\cA\to\C\quad ,\quad \cA^+\to \R^+ \;\text{with}\; \omega(\mbf{1})=1  \eeq
Its physical role is that of an expectation functional. With the help of the ordinary Hilbert space trace, $tr$, and a density matrix $\rho\in\cA$ and $tr(\rho)=1$ we can write a state as
\beq  \omega(A)=tr(\rho\cdot\cA)  \eeq

 In a type $II$ v.N. algebra we have no minimal projections, hence no pure states. Furthermore, the ordinary trace, $tr$, is infinite on the projectors and of no use. Henced, in a first step, we have to introduce a generalized concept, denoted by $Tr$.
\begin{defi}A (faithful) trace, $Tr$, on the v.N.algebra , $\cA$, is a map from $\cA^+$ to $[0,\infty]$ with:\\
\begin{align} i) Tr(\lambda A)&=\lambda A\;\text{for}\; A\in\cA^+\;\text{and}\;\lambda\geq 0  \\
ii) Tr(A+B)&=Tr(A)+Tr(b)\;\text{for}\; A,B\in \cA^+ \\
iii) Tr(UAU^*)&= Tr(A)\;\text{for}\; A\in\cA^+\, U\,\text{unitary in}\,\cA \\
iv) Tr(A)&=0\;\text{implies}\; A=0 
\end{align}
\end{defi}
\begin{bem}For the sake of brevity we leave out some further technical assumptions (e.g. continuity aspects) and refer to the above cited literature and, in particular, to the concise discussion in \cite{Sorce}. Furthermore, while physicists are presumably more accustomed to the above numerical trace, in most of the modern textbooks the socalled center-valued trace is introduced. For factors, however, the two notions are the same.
\end{bem}

Some of the various useful properties of this notion, which can be derived from the above assumptions are for example:
\begin{satz}It holds\\
i) $Tr(AA^*)=Tr(A^*A)$\\
ii)The operators $A\in\cA$ with $Tr(|A|)<\infty$ form an ideal of the trace-class operators. $Tr$ can be linearly extended to this ideal $\cA_1$.\\
iii)We have for $A\in\cA_1,B\in \cA:Tr(AB)=Tr(BA)$\\
iv)For $P,Q$ projectors in $\cA$ it holds:\\
$P\sim Q\leftrightarrow Tr(P)=Tr(Q)\;,\,P\prec Q\leftrightarrow Tr(P)<Tr(Q)$
\end{satz}
\begin{lemma}From the above it follows that with $P$ infinite we have $Tr(P)=\infty$.
\end{lemma}
Proof:$P$ infinite implies, there does exist a $Q$ with $Q<P$ but $Q\sim P$. Hence 
\beq Tr(P)=Tr(Q+(P-Q))=Tr(Q)+Tr(P-Q)=Tr(P)+Tr(P-Q)  \eeq
with $Tr(P-Q)\neq 0$, which is a contradiction.

The interesting case in our context are v.N. algebras of type $II_1$. The following can be shown (non-trivial):
\begin{satz}Therer exists a faithful finite unique tace, $Tr$, up to rescaling, in a type $II_1$ v.N. algebra.
\end{satz}
Now we have everything we need to infer useful information from this quite abstract framework. First of all, with the help of this unique trace, $Tr$, we can calculate generalized expectation values of observables $A\in\cA$ ($\cA$ of type $II_1$), i.e. $Tr(A)$. Furthermore, with $\rho\in\cA^+$, we can calculate (thermal) averages for $B\in\cA$:
\beq <B>:=Tr(\rho B)\;,\; Tr(\rho):=1  \eeq
i.e., with $\rho$ playing the role of a density matrix.

What is, in fact, remarkable is the following: We have the class of finite projections in $\cA^+$. As we have no minimal projections in a type $II_1$ v.N. algebra, that is, a fortiori no pure states, the admissible projectors project on, in effect, infinite dimensional subspaces. The spectral theory of the observables in $\cA$ imply that in the spectral resolution of $A\in\cA$ only these projectors with higher dimensional rank can occur.
\begin{conclusion}In contrast to the type $I$ situation of e.g. $\cB(\cH)$, the possible observables which can occur in a type $II_1$ algebra are of a rather coarse grained character. The same holds for the possible preparation of states. This points in the direction of a (quasi) thermal character of the kind of physics under discussion (see below).
\end{conclusion}
What we have said above complies well with a remark in the introduction of \cite{W1}.
\begin{quote}\ldots type $II$ means that we do not understand the microstates of a BH but we do have a framework to analyze the BH entropy.
\end{quote}

We now want to relate the above results with two strands of our own investigations in quantum gravity (QG) and quantum space-time physics (QST). In \cite{Requ2} we argued that the substructure of QG and QST is of a thermal character. In \cite{Requ1} we developed a theory of the microstructure of QST and the singularity of a BH. On the other hand, in \cite{Requ5} we introduced what we call a \tit{Geometric Renormalization Group}, a framework which started from quite erratic discrete structures (essentially arrays or networks of microscopic degrees of freedom (DoF) and their fluctuating or stochastically distributed network of elementary interactions or entanglement relations (as to the latter topic see also \cite{Requ1} sect. 3.2). We employed highly advanced mathematics, developed by, for example, M.Gromov (a metric superspace with its elements being metric spaces) to study a possible continuum limit of such erratic spaces and the emergent limit properties.

We combined this mathematical structure with our own approach (see e.g. \cite{Requ4} or \cite{Requ3}) to construct consecutive levels of coarse graining in QST, connected vertically by well-defined coarse graining maps (essentially being quasi isometries) and horizontally by the respective continuum limit constructions (essentially being geometric scaling maps).
\begin{conclusion}We think that these coarse graining levels of QST and their respective continuum limits are closely related to the structure of type $II_{\infty}$ v.N. algebras and their tensor structure of an exterior ordinary Hilbert space algebra $\cB(\cH_I)$ with an internal algebra of type $II_1$.
\end{conclusion}
\begin{bem}Quantum mechanically the reöation of the macroscopic points, $x$, living on the smooth space-time manifold $ST$ to these infinitesimal neighborhoods, $[x]$, resemble the relation of the macro observables of a macroscopic quantum system to the microscopic DoF of the same quantum many body system.
\end{bem}

In \cite{Requ2} and \cite{Requ1} we studied the thermal substructure of QST and GR. we argued that the metric tensor, $g_{ij}(x)$, is kind of an \tit{order parameter field} on (Q)ST or ST and that, as $[x]$ contains many gravitational DoF, it represents with high probability a thermal system. This is, among other things, a consequence of the interesting phenomenon of \tit{typicality} (one of the early inventors being v.Neumann).

If we now put together what we have said above and relate it to the type theory of $II_{\infty}$ v.N. algebras, we will formulate an observation and a conjecture.
\begin{ob}The resolution of the type $II_{\infty}$ v.N. algebra $\cA$ into a sum $\sum_{ij=0}^{\infty}P_i\cA P_j$ with $P_i$ mutually orthogonal finite projectors and $\sum\,P_i\cH=\cH$ shows that the underlying quantum system can be thought of as consisting of a countable infinity of (local) subsystems which are interacting with each other. These subsystems, $P_i\cA P_i$, are of type $II_1$ and the subprojectors they contain have infinite rank. Therefore they admit only a relatively coarse spectral resolution of the observables lying in $P_i\cA P_i$ of (possibly) a thermal character. Depending on the choice of $P_0$ or $\{P_i\}$ we may have different levels of  coarseness of observations or preparations.
 \end{ob}
\begin{conjecture}We provide the following physical interpretation of the structure $\cA\cong\cB(\cH_I)\overline{\otimes} P_0\cA P_0$. On the surface level of QST we have QFT on curved space-time ST as the socalled semiclassical approximation, i.e., gravitation treated classically. While we do not think that quantum theory necessarily holds all the way down to the most primordial levels, we think it holds at least on the coarse grained levels next to the smooth surface level of the scenario we have described above. We then conjecture that the above tensor product structure describes the first steps away from the semiclassical approximation in the direction of a full theory of quantum gravity . That is, the ordinary Hilbert space structure $\cH_I$ and its full algebra $\cB(\cH_I)$ describes ordinary quantum matter, $QM$, interacting locally around each macroscopic point, $x$, with the underlying gravitational DoF, encoded in $P_i\cA P_i$ or $P_0\cA P_0$. 
\end{conjecture}

\end{document}